\documentclass[preprint,aps,prc,superscriptaddress,showpacs,floatfix]{revtex4}
\usepackage{graphicx}

\newcommand{\nc}{\newcommand}
\nc{\pt}{p_{\rm T}}
\nc{\ov}{\overline}
\nc{\beq}[1]{\begin{equation}\label{#1}}
\nc{\eeq}{\end{equation}}
\nc{\bea}[1]{\begin{eqnarray}\label{#1}}
\nc{\eea}{\end{eqnarray}}

\begin{document}

\title{Effect of space-momentum correlations on the constituent quark 
number scaling of hadron elliptic flows} 

\author{V. Greco}
\affiliation{Cyclotron Institute and Physics Department, Texas A\&M 
University, College Station, Texas 77843-3366}
\author{C. M. Ko}
\affiliation{Cyclotron Institute and Physics Department, Texas A\&M 
University, College Station, Texas 77843-3366}

\date{\today}

\begin{abstract}
Using models ranging from schematic one with a simple quark
distribution to more realistic blast wave, we study the 
elliptic flow of hadrons produced from coalescence of quarks and
antiquarks in the quark-gluon plasma that is formed in 
ultrarelativistic heavy ion collisions. In particular, we 
study effects due to azimuthal anisotropy in the local
transverse momentum distribution of quarks, as generated by their
position-momentum correlations as a result of radial flow 
and/or jet quenching. We find that even if quarks have large
local non-elliptic anisotropic flow, the elliptic flow of produced 
hadrons can still scale with their constituent quark numbers.  This 
scaling is, however, broken if the radial flow of coalescing quarks  
is anisotropic and/or if the momentum dispersion of quarks inside 
hadrons is included. 
\end{abstract}

\maketitle

\section{introduction}
In searching for the signatures of the Quark-Gluon Plasma (QGP) produced 
at the Relativistic Heavy Ion Collider (RHIC) and to study its properties, 
the elliptic flow, i.e., the second harmonic in the azimuthal distribution
of the transverse momenta of produced particles, has played an important 
role.  It was shown in Ref.\cite{zhang} using the parton cascade 
model and in Refs.\cite{tean,kolb} using the hydrodynamic model
that the elliptic flow in heavy ion collisions at RHIC is sensitive to
the properties of formed partonic matter. The observed large elliptic 
flow of charged hadrons at RHIC \cite{adlv2,adams} is indeed consistent 
with an initial partonic matter that is not only dense but also strongly 
interacting \cite{shuryak04}. For identified midrapidity hadrons except 
pions, it was further observed that the dependence of the elliptic flow 
on transverse momentum becomes similar if both are divided 
by the number of constituent quarks in the hadrons, i.e., two for 
mesons and three for baryons. Such scaling of the elliptic flow
of hadrons according to their constituent quark numbers 
\cite{sore,voloshin} is consistent with the picture of constituent 
or massive quarks coalescing into hadrons 
\cite{moln-v2,kolb1,fries2,greco2,hwa}. In particular, the 
quark number scaling of hadron elliptic flow is exact in the so-called 
naive coalescence model, which only allows quarks and antiquarks 
(generally called quarks in the rest of the paper) with same 
transverse momentum to coalescence into hadrons. This scaling
is, however, violated in more realistic quark coalescence models 
that takes into account the momentum distribution of quarks inside hadrons 
\cite{fries2,greco2,lin-moln,greco-res} or higher Fock states in 
hadron wave functions \cite{muller}. According to Ref.\cite{greco-res}, 
the former leads to a violation of the quark number scaling of hadron
elliptic flow by about 10\% for mesons and 15\% for baryons, resulting 
in a relative baryon/meson violation of about 5\% which is similar to 
that observed in the most recent experimental results 
\cite{sorensen-qm05}. The hadron elliptic flow is also affected by 
resonance decays \cite{greco-res}. Although this effect has been estimated 
to be small, including the resonance decay effect can largely account 
for the observed quark number scaling violation in the pion elliptic 
flow \cite{greco-res,dong}.

Other sources for the violation of the quark number scaling in 
hadron elliptic flow have also been studied recently.  For example, 
it was shown that anisotropies in the quark phase-space density 
and effective emission volume at hadronization can lead to the 
breaking of hadron elliptic flow according to their constituent 
quark numbers \cite{pratt}. Furthermore, it was pointed out 
in Ref. \cite{moln-v1} that in usual derivation of the quark 
number scaling of the elliptic flow based on the naive 
quark coalescence model \cite{moln-v2,fries2,kolb1}, correlations 
between positions and momenta of quarks, as expected from the
hydrodynamic or the parton cascade model, were ignored as 
only the spatially averaged quark momentum distributions
were used. Since the position-momentum correlations can generate 
large non-elliptic anisotropy in the quark local momentum distribution, 
which may not be averaged out after integrating over the quark spatial 
distribution, the quark number scaling of hadron elliptic flows may 
thus be strongly violated.  

To understand more clearly what role is played by quark 
position-momentum correlations and how the quark number scaling 
of hadron elliptic flow appears in the quark coalescence model, we 
examine in the present paper the relation between the elliptic flows
of quarks and mesons in models that range from a simple schematic model 
to the more realistic blast wave model. Our analyses indicate that
although large azimuthal anisotropy generally exists in the 
quark local transverse momentum distributions in these diverse models as 
well as in the model used in Ref.\cite{greco2}, it does not necessarily 
lead to the violation of the quark number scaling of hadron elliptic flows 
if the quark global transverse momentum distribution does not have 
odd-order anisotropic flows. 

\section{Anisotropic flow}\label{flow}

Following the notations of Ref.\cite{moln-v1}, the spatial and transverse
momentum distributions $S(x,{\bf p}_T, y)$ of particles produced in 
relativistic heavy ion collisions can be expanded in Fourier series as 
\beq{expansion}
S(x,{\bf p}_T, y)= S_0(x,{\bf p}_T, y)\left \{ 1+ 2 \sum_{n=1}^\infty 
{\rm Re}\left( c_n (x,p_T,y) e^{-in\phi_p}\right)\right \},
\eeq
where ${\bf p}_T \equiv p_T (\cos \phi_p, \sin \phi_p)$ is the
transverse momentum of a particle with $\phi_p$ denoting its azimuthal
angle with respect to the reaction plane, and $y$ is its rapidity. 
The Fourier coefficient $c_n(x,p_T,y)$ denotes the $n$th-order 
{\it local} anisotropic flow of particles with momentum $p_T$ in an 
infinitesimal space-time volume around $x$. In general, $c_n$ is  
complex and is written as $c_n = v_n + iu_n$, with the real and imaginary 
parts given by averaging over the azimuthal angle $\phi_p$ of 
particle momentum, i.e., $v_n = \langle \cos (n\phi_p) \rangle$ and 
$u_n = \langle \sin (n\phi_p) \rangle$. 

Experimentally, only the spatially averaged global anisotropic flows are 
measured, i.e., 
\beq{cn-av}
\ov{c}_n (p_T,y)\equiv\langle c_n(x,p_T,y)\rangle= 
\frac{\int d^4 x c_n(x,p_T,y)S(x,{\bf p}_T,y)}{\int d^4 x S(x,{\bf p}_T,y)}.
\eeq
The real part of the two lowest orders $\ov{v}_1$ and $\ov{v}_2$ are
the directed and elliptic flows, respectively. The imaginary part 
$\ov u_n$ always vanish in collisions between identical spherical nuclei 
due to reflection symmetry with respect to the reaction plane, i.e.,  
invariant under the transformation $\phi \rightarrow - \phi$. 
Furthermore, for particles at midrapidity in collisions with equal 
mass nuclei, anisotropic flows of odd orders vanish as a result of the
additional symmetry $\phi \rightarrow \phi+\pi$. In the present study, 
we consider only particles at midrapidity in collisions of identical
spherical nuclei. 

In the naive coalescence model \cite{moln-v2,fries2,kolb1}, which 
only considers the momentum distribution of quarks, thus neglecting 
correlations between quark positions and momenta, and allows 
quarks with same transverse momentum to coalesce into hadrons, one 
obtains a simple relation between meson and quark elliptic 
flow if all $\ov v_n$ of quarks vanish except for $n=2$; i.e, 
\beq{v2_scal_mes}
\ov v_{2,M} = \frac{2{\ov v}_{2,q}}{1 + 2 {\ov v}^2_{2,q} }.
\eeq
Neglecting ${\ov v}^2_{2,q}$ in the denominators due to its small 
value, a scaling of the elliptic flow of mesons according to their 
constituent quark number then follows.  

Including quark position-momentum correlations, the meson local 
elliptic flow $v_{2,M}$ is then related to quark local $v_n$ and 
$u_n$ via 
\cite{moln-v1}
\beq{vn_general}
v_{2,M}=\frac {2 v_2 + v_1^2-u_1^2+2\sum_{k=1}^{\infty}
(v_k v_{2+k}+u_k u_{2+k})}{1+2 \sum_{k=1}^{\infty} (v_k^2+u_k^2)}.
\eeq
Although higher-order quark local anisotropic flows $v_n$ and $u_n$
with $n>3$ are likely small, quark local $v_1$ and $u_1$ can be large
as first pointed out in Ref.\cite{moln-v1}.  It is not obvious 
that in the spatial average the contribution of quark local directed
flow $v_1$, $u_1$ vanishes. 
The observed approximate scaling of hadron elliptic flow
according to the constituent quark number may simply due to accidental 
cancellations of quark local directed flow as suggested in 
Ref.\cite{moln-v1}. In the present paper, we study the effect of 
nonvanishing quark local directed flow on the meson elliptic flow in 
various models. Our analysis shows that after spatial average
its effect either disappears or is small, thus having negligible 
or only small effect on the elliptic flow of hadrons. 

\section{A schematic model}\label{schematic}

We first consider a schematic model in which quarks are distributed 
uniformly in space with their momenta pointing radially outward, 
i.e., same azimuthal angles $\phi_p$ and $\phi_r$ for the 
momentum and position vectors of a quark. This quark distribution 
is qualitatively similar to that of quenched jets produced in heavy ion 
collisions  at RHIC \cite{gyulassy,dainese}. Including also an azimuthal 
anisotropy of the form $\cos(2 \phi_p)$ in the quark transverse
momentum distribution, the quark distribution becomes 
\beq{qdistrb1}
\frac{dN}{d^2{\bf r}d^2{\bf p}}\propto\left(1+2 a_2 \cos(2 \phi_p)\right) 
\delta(\phi_r -\phi_p),
\eeq
where the coefficient $a_2$ is the quark elliptic flow $\ov v_{2,q}$. 

Considering only quarks with spatial azimuthal angles less than 
$\phi_r^\prime$, the Fourier coefficient $c_{n,q}$ of resulting 
momentum distribution can be evaluated by integrating $\phi_r$ from 0 to
$\phi_r^\prime$, i.e., 
\bea{vnq-distrb1-expl}
c_{n,q}&=&\frac{1}{N}
\int_0^{\phi'_r} d\phi_r \int_0^{2 \pi} d\phi_p \left[1+ 2 a_2 
\cos (2 \phi_p)\right]\delta(\phi_r -\phi_p)e^{in \phi_p}\nonumber\\
&=&\frac{i}{N}\left\{\frac{1}{n} + a_2 \frac{2n}{n^2-4} 
-\frac{e^{i n\phi'_r}}{n} - 2 a_2 
\left[\frac{e^{i (n-2)\phi'_r}}{n-2} + \frac{e^{i (n+2)\phi'_r}}{n+2}
\right] \right\}, 
\eea
where $N= \phi'_r + a_2 \sin (2 \phi'_r)$ is the normalization factor
shown in the denominator of Eq.(\ref{cn-av}). For $n=2$, the above 
equation gives 
\beq{v2q-distrb1-expl}
c_{2,q}=\frac{1}{N}\left\{a_2 \phi'_r + i \left[\frac{1}{2} 
+ \frac{a_2}{4} - \left(\frac{1}{2} + \frac{a_2}{4} e^{2i
\phi'_r }\right)e^{2i \phi'_r }\right]\right\}.
\eeq

The global quark elliptic flow is obtained from Eq.(\ref{v2q-distrb1-expl})
by letting $\phi'_r=2\pi$, and this gives $\overline {v}_{2,q} = a_2$ 
as expected. It can be checked that all higher-order global quark 
anisotropic flows vanish as the first two terms in Eq.(\ref{vnq-distrb1-expl}) 
are exactly canceled by the last two terms when $\phi'_r=2\pi$.  On 
the other hand, for quarks in a localized space the distribution 
contains an infinite number of Fourier coefficients. For example, for 
quarks in the first quadrant, which corresponds to integrating $\phi_r$ 
from 0 to $\phi'_r = \pi/2$, all $c_n$'s are nonzero. Assuming an 
anisotropy parameter $a_2=0.1$, we find that for quarks in the first 
quadrant, the local directed flow is $v_1=0.68$ and its 
imaginary part is $u_1=0.59$. Both are close to those from the MPC 
code for Au+Au collisions at $\sqrt{s}= 200$ GeV with $b=8$ fm 
\cite{moln-v1}. Except for the local elliptic flow 
$v_{2,q}=\overline{v}_{2,q}=0.1$, other real even Fourier coefficients 
all vanish, and this is similar to that in the midrapidity region 
of heavy ion collisions at RHIC. The imaginary even Fourier 
coefficients are in general nonzero, and we have for the second 
Fourier coefficient $u_{2,q}=0.64$. We note that the large azimuthal 
anisotropy in the quark local momentum distribution in the present 
model is caused by the strong position and momentum correlations in the 
quark distribution, which is also present in the blast wave model 
\cite{greco} as well as in the transport model \cite{moln-nscal}. 
These models further include, however, effects due 
to thermal motions of quarks. 

In the naive coalescence model, the transverse momentum spectrum 
of produced mesons is given by \cite{greco2}
\bea{coal2}
\frac{dN_M}{d^2{\bf p}}= g_M \int{\prod_{i=1}^{2} d^2{\bf r}_i 
d^2{\bf p}_i\frac{dN_q}{d^2{\bf r}_i d^2{\bf p}_i}
(2\pi)^2\delta^{(2)}({\bf r}_1-{\bf r}_2) 
\delta^{(2)}({\bf p}_1-{\bf p}_2)\delta^{(2)}({\bf p}-{\bf p}_1-{\bf p}_2)},
\eea
where $g_M$ is the statistical factor for a quark and an antiquark to 
form a colorless meson with certain spin and isospin. 

With the quark distribution of Eq.(\ref{qdistrb1}), carrying out 
the integral in Eq.(\ref{coal2}) over the azimuthal angle from 
0 to $\phi_r^\prime$ gives following meson anisotropic flow 
in a restricted space: 
\bea{vnm-distrb1-expl}
c_{n,M}=&&\frac{i}{N}\Bigg\{\frac{1+2a_2^2}{n}+\frac{4na_2}{n^2-4}
+\frac{2na_2^2}{n^2-16}-\frac{1+2a_2^2}{n}e^{i n\phi'_r}\nonumber\\
&&-2 a_2\left[\frac{e^{i(n-2)\phi'_r}}{n-2}+\frac{e^{i (n+2)\phi'_r}}{n+2}
\right]-a_2^2\left[\frac{e^{i(n-4)\phi'_r}}{n-4}
+\frac{e^{i(n+4)\phi'_r}}{n+4}\right]\Bigg\},
\eea
where $N=(1+2a_2^2)\phi'_r+2a_2\sin (2\phi'_r)+a_2^2/2 \sin(4\phi'_r)$.
The local elliptic flow in this restricted space is then
\bea{v2m-distrb1-expl}
c_{2,M}=&&\frac{1}{N}\left\{2 a_2 \phi'_r+ i\left[\frac{1+2a_2^2}{2} 
+\frac{a_2}{2}-\frac{2a_2^2}{3}-\frac{1+2a_2^2}{2}e^{i 2\phi'_r}  
\right.\right.\nonumber\\
&&\left.\left.-a_2\frac{e^{4 i\phi'_r}}{2}
-a_2^2\left(\frac{e^{6 i\phi'_r}}{6}-\frac{e^{-2 i\phi'_r}}{2}\right)
\right]\right\}.
\eea

Including all quarks, i.e., letting $\phi_r^\prime=2\pi$ in 
Eq.(\ref{vnm-distrb1-expl}), cancellations similar to those in the 
evaluation of the quark global anisotropy flows lead 
to the following meson transverse momentum distribution:
\bea{coal2-expl}
\frac{dN_M}{d^2{\bf p}}\propto 1+2a^2_2+ 4a_2\cos(2\phi_p)
+ 2a^2_2\cos(4\phi_p).
\eea
The above equation shows that the meson elliptic flow is  
\begin{equation}
\ov v_{2,M} = \frac{2{\ov v}_{2,q}}{1 + 2 {\ov v}^2_{2,q} },
\end{equation}
and there is also a 4th-order anisotropic flow given by 
\begin{equation}
{\ov v}_{4,M}=\frac{{\ov v}^2_{2,q}}{1 + 2{\ov v}^2_{2,q} }.
\end{equation}
These results are exactly what one expects from the naive
coalescence model when all quark anisotropic flows $\ov c_n$ 
except $\ov v_2$ are zero \cite{kolb1}. 

The above result holds for any value of the anisotropy parameter $a_2=\ov
v_{2,q}$, and thus for any value of local anisotropic flow $c_{n,q}$.
In other words, despite the fact that locally there can be very large
anisotropy in the quark momentum distribution, they are not
relevant for the global elliptic flow of mesons that are produced 
from the coalescence process. The quark number scaling of hadron elliptic
flow is exact as expected when the quark distribution has only non-zero 
global elliptic flow. 

\section{The blast wave model}

The schematic model introduced above can be made more realistic by  
including thermal motions of quarks. Adding a random component to 
the quark momentum weakens the extreme angular correlation between 
the positions and momenta of quarks in the schematic model. 
As shown in the following, mesons produced from the naive coalescence 
remain to have an elliptic flow that follows the quark number scaling 
if the radial flow is isotropic, i.e. independent of azimuthal angle. 
The large anisotropy in the quark local momentum distribution, 
particularly the $v_{1,q}$ and $u_{1,q}$, does not affect the scaling 
relation between quark and meson elliptic flows. On the other hand, 
the quark number scaling of hadron elliptic flow can be violated when 
the quark radial flow is anisotropic, i.e., its velocity varies with 
the spatial azimuthal angle $\phi_r$. 

\subsection{Blast wave with isotropic radial flow}\label{blast}

We first consider the case that quarks have a relativistic
Boltzmann distribution with isotropic radial flow velocity 
\cite{heinz}, i.e.,  
\bea{blast1-q}
\frac{dN_{\rm q,\bar q}}{d^2{\bf p}_Td^2{\bf r}}
\propto\exp\left[-\frac{m_T}{T}\cosh\rho_0
+\frac{p_T}{T}\sinh\rho_0 \cos(\phi_r-\phi_p)\right] 
\left[1+2 a_2(p_T)\cos(2\phi_p) \right],
\eea
where $m_T=(p_T^2+m_q^2)^{1/2}$. In the above, dependence on the spatial
and momentum rapidities $y$ and $\eta$ has been dropped as we consider 
a Bjorken boost invariant distribution at $y=\eta=0$.  For simplicity, 
both the quark density distribution and their flow profile 
$\beta_T=\tanh\rho_0$ are taken to be independent of the radial 
position. Also, the spatial shape of the quark distribution is 
assumed to be cylindrical. Although spatial ellipticity can cause 
the breaking of quark number scaling of hadron elliptic flow 
as discussed in Ref.\cite{pratt}, it is not essential for the 
purpose of present study. 

For quarks with spatial azimuthal angle less than $\phi_r^\prime$,
the different harmonics in the Fourier expansion of the quark
distribution in Eq.(\ref{blast1-q}) are given by
\bea{cnq-blast1}
c_{n,q}(p_T)=&&\frac{1}{N}\int_{0}^{2\pi} d\phi_p 
\int_{0}^{\phi'_r}d\phi_r\exp\left[\alpha(p_T)\,\cos(\phi_r-\phi_p)
+i n\phi_p\right]\nonumber\\
&&\times\left[1+2\,a_2(p_T)\cos(2\phi_p) \right],
\eea
with
$N=2\pi\{I_0[\alpha(p_T)]\phi'_r+a_2(p_T)I_2[\alpha(p_T)] 
\sin(2\phi'_r)\}$ and $\alpha(p_T)=(p_T/T)\sinh \rho_0$.
If all quarks are included, i.e., $\phi_r^\prime=2\pi$, then
all orders of anisotropic flow except $\ov{v}_2$ vanish. This can be 
easily seen as the spatial integral can now be expressed in terms of 
the Bessel function $I_0$ and is thus canceled by the same factor in the 
normalization factor $N$ in the denominator. What is left is then 
the integral over $\phi_p$ which is non-zero only for $n=2$. 

\begin{figure}[ht]
\includegraphics[height=2.7in,width=3.5in]{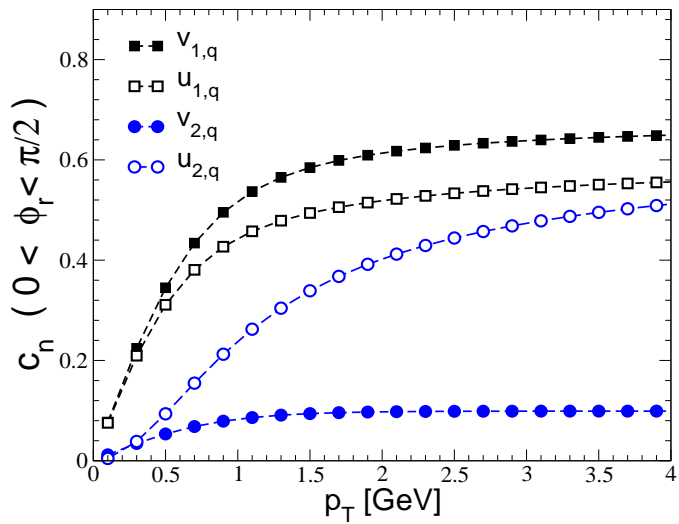}
\caption{Quark local anisotropic flow as a function of transverse 
momentum from averaging over quarks in the first quadrant of the 
transverse plane.}
\label{cnq-blastv0}
\end{figure}

On the other hand, because of quark radial flow, none of the
harmonics in the Fourier expansion vanishes if only quarks in a 
limited space are considered.  To estimate the values of 
the different harmonics in the present model for heavy ion 
collisions at RHIC, we take a slope parameter $T=170$ MeV for the
quark transverse momentum distribution, which is consistent with the 
phase transition temperature from lattice QCD calculations.  Masses 
of quarks are taken to be $m_q = m_{\bar q}=300$ MeV as in 
Ref.\cite{greco2}. For the strength of quark anisotropic flow, we
take $\rho_0 =0.4$ and $a_2(p_T)= 0.1\tanh(1.2p_T)$ in order to 
have radial and elliptic flows with magnitude similar to those 
observed in experiments. Results for the first two local 
harmonics $v_{1,q}$, $u_{1,q}$, $v_{2,q}$ and $u_{2,q}$ in the
first quadrant are shown in Fig.\ref{cnq-blastv0}, and they are seen 
to have values similar to those from the MPC quark cascade model 
\cite{moln-nscal}.

The meson spectrum obtained from the naive coalescence of quarks, 
i.e., only quarks in the same phase space point can coalesce, can be
easily derived from the quark distribution of Eq.(\ref{blast1-q})
by integrating over both spatial azimuthal angle $\phi_r$ and momentum
azimuthal angle $\phi_p$, and the result is given by 
\bea{blastv2-q}
\frac{dN_M}{d^2{\bf p}_T}\propto&&
e^{- 2 \tilde\alpha(p_T/2)} I_0\left[2\alpha(p_T/2) \right]\nonumber\\
&&\times\left[1+ 2a_2^2(p_T)+ 4 a_2(p_T)\cos(2\phi_p)
+2 a_2^2(p_T) \cos(4\phi_p) \right],
\eea
where $\tilde{\alpha}(p_T)=(m_T/T)\cosh\rho_0$. Comparing to 
Eq.(\ref{coal2-expl}), we see that the quark number
scaling of meson elliptic flow appears as in the schematic 
model of Section \ref{schematic}. This example thus demonstrates again 
that the naive coalescence model leads to the quark number scaling of 
hadron elliptic flow, independent of the presence of large
anisotropy in the quark local momentum distribution. 

If we had considered quarks in a finite rapidity range, integrating
over the rapidity leads to the Bessel function $K_1(\tilde\alpha)$
instead of the $e^{-\tilde \alpha}$ in Eq.(\ref{blastv2-q}). 
This does not, however, affect the conclusion from the simpler picture
considered in the above.

\subsection{Blast wave with anisotropic radial flow}\label{ablast}

The above blast wave model can be made more general by allowing the
radial flow velocity to depend on both the azimuthal
angle in space and the transverse momentum. Specifically, we 
introduce a radial flow profile 
$\beta_T(p_T,\phi_r)=\tanh\rho(p_T,\phi_r)$, with 
$\rho(p_T,\phi_r) =\rho_0+\rho_2(p_T)\cos(2\phi_r)$.
This is similar to that in the blast wave model of
Ref.\cite{huovinen}, except for the $p_T$ dependence of the azimuthal 
anisotropy parameter $\rho_2$. The latter makes it possible to 
parameterize the typical increase and then saturation (or slight decrease) 
of the elliptic flow with transverse momentum, that is observed in 
experiments \cite{adlv2,adams} and predicted by transport models 
\cite{linko,chen,molnar}. With anisotropic radial flow, quarks can 
acquire an elliptic flow without introducing explicitly anisotropic  
azimuthal distributions in space and momentum as in the models 
discussed so far. The quark distribution in the present model is 
then given by 
\bea{blast0-q}
\frac{dN_{\rm q,\bar q}}{d^2{\bf p}_Td^2{\bf r}}\propto
\exp \left[- \frac{m_T}{T}\cosh\rho(p_T, \phi_r)
+\frac{p_T}{T} \sinh\rho(p_T,\phi_r) \cos(\phi_r-\phi_p)\right].
\eea

Defining $\tilde\alpha(p_T,\phi_r)=(m_T/T)\cosh \rho(p_T,\phi_r)$ and
${\alpha}(p_T,\phi_r)=(p_T/T)\sinh\rho(p_T,\phi_r)$, the transverse 
momentum dependence of the quark elliptic flow can be easily
calculated as in Ref.\cite{huovinen}, and the result 
is given by
\beq{v2q-blast}
{\ov v}_{2,q}(p_T)= \frac{\int_{0}^{2\pi}d\phi_r \exp 
\left[- \tilde\alpha(p_T, \phi_r)\right]
I_2\left[\alpha(p_T,\phi_r)\right]\cos(2\phi_r)}
{\int_{0}^{2\pi}d\phi_r \exp \left[- \tilde\alpha(p_T, \phi_r)\right]
I_0\left[\alpha(p_T,\phi_r)\right]}.
\eeq

In the naive quark coalescence model, which only allows quarks with 
same momentum to coalesce into hadrons, the meson distribution is 
simply given by the following one-dimensional integral:
\beq{dndpm-blast}
\frac{dN_{M}}{d^2{\bf p}}\propto\int_{0}^{2\pi}d\phi_r\exp 
\left[- 2\tilde\alpha(p_T/2, \phi_r)\right]
\exp \left[2\alpha(p_T/2,\phi_r)\cos(\phi_r-\phi_p)\right].
\eeq
The elliptic flow of mesons is then 
\beq{v2m-blast}
{\ov v}_{2,M}(p_T)= \frac{\int_{0}^{2\pi}{d\phi_r\exp 
\left[-2\tilde\alpha(p_T/2, \phi_r)\right]
I_2\left[2\alpha(p_T/2,\phi_r)\right]\cos(2\phi_r)}}
{\int_{0}^{2\pi}d\phi_r\exp\left[-2\tilde\alpha(p_T/2, \phi_r)\right]
I_0\left[2\alpha(p_T/2,\phi_r)\right]}.
\eeq
Contrary to the case of isotropic radial flow, the quark number scaling of
hadron elliptic flow is not obvious when the radial flow is 
anisotropic, and a violation is generally found as shown below.

\begin{figure}[ht]
\includegraphics[height=2.7in,width=3.5in]{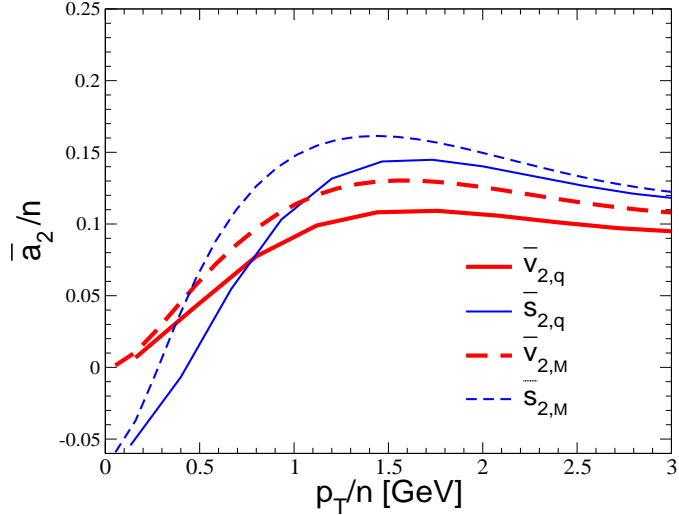}
\caption{Scaled momentum elliptic flow $\ov{v}_2$ and spatial elliptic
deformation $\ov{s}_2$ as a function of scaled transverse momentum.
Solid lines are the quark $\ov{v}_{2,q}$ (thick) and $\ov{s}_{2,q}$ 
(thin), while dashed lines are $\ov{v}_{2,M}$ (thick) and 
$\ov{s}_{2,M}$ (thin) of mesons.}
\label{v2qm-blast}
\end{figure}

Using the anisotropic function
$\rho_2=0.25/\{1+\exp(p_T-0.5)/0.75]\}+0.014$ in the quark radial flow, 
we have evaluated numerically both the quark and scaled meson elliptic 
flows, and they are shown in Fig.\ref{v2qm-blast} by the solid 
and dashed thick lines, respectively. It is seen that the quark number 
scaling of hadron elliptic flow is appreciably broken (about $20\%$) 
at $p_T/n \sim$ 1.5 GeV. Furthermore, the meson elliptic flow in the 
present model is larger than the one expected from the quark number 
scaling, in contrast with the smaller meson elliptic flow from most 
other sources of quark number scaling violation 
\cite{lin-moln,greco-res,pratt,muller}. Since the local harmonics 
in the quark momentum distribution in the present case have values 
very similar to those in the previous blast wave model with isotropic
radial flow, they are not the reason that the quark number 
scaling of meson flow is violated. The violation is caused instead by 
the larger coalescence probability for quarks in the reaction plane 
than out of the reaction plane as a result of anisotropy in the 
radial flow. This can be seen by examining the second harmonic in the 
spatial distribution $s_2 =\langle \cos(2\phi_r) \rangle$ of quarks 
and mesons. As shown in Fig.\ref{v2qm-blast}, the scaled spatial 
elliptic deformation $s_2$ of mesons (thin dashed line) is indeed 
larger than that of the quarks (thin solid line), except at low
transverse momenta. 

Our results at low transverse momenta needs, however, to be interpreted 
with caution as the coalescence model violates unitary if applied to 
the entire phase space. Although the coalescence model is applicable 
for describing the production of rare particles such as those with 
intermediate transverse momenta, a more dynamical approach that 
takes into account properly energy and entropy conservation is 
needed for more abundant low transverse momenta particles. 

We would like to point out that for quarks with transverse momentum 
$p_T/n\geq 1.5$ GeV, their elliptic flow is not likely affected 
by anisotropy in their radial flow. The momentum distribution of these
quarks is thus more similar to that in the schematic model of  
Sect. \ref{schematic} for which the quark number scaling of hadron 
elliptic flow holds exactly.

\section{Hadron wave function effect on elliptic flow}

Besides anisotropic radial flow, the quark number scaling of
hadron elliptic flow can also be violated by the coalescence of 
quarks with different momenta and at different positions when the hadron
wave function is taken into consideration as in Refs.\cite{greco2,greco-res}.
To see this explicitly, we generalize Eq.(\ref{coal2}) to three
dimensional space and replace the resulting two delta functions 
$(2\pi)^3\delta^{(3)}({\bf r}_1-{\bf r}_2)\delta^{(3)}({\bf p}_1-{\bf p}_2)$ 
by the meson Wigner function from Ref.\cite{greco2}, i.e., 
\beq{wigner}
f_M({\bf x}_1,{\bf x}_2;{\bf p}_1,{\bf p}_2)=\frac{9 \pi}{2}
\Theta\left(\Delta_x^2-({\bf x}_1 - {\bf x}_2)^2\right)
\Theta\left(\Delta_p^2-({\bf p}_1 - {\bf p}_2)^2\right),
\eeq
where $\Delta_x = \Delta_p^{-1}$ is the width of the quark momentum 
distribution in the meson.

\begin{figure}[ht]
\includegraphics[height=4in,width=3.5in,angle=-90]{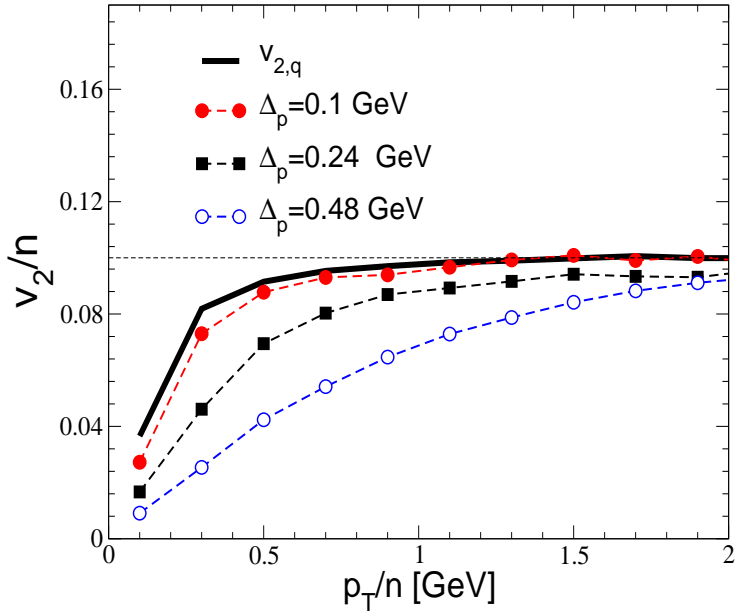}
\caption{Scaled elliptic flow as a function of scaled transverse 
momentum. The solid line is the quark elliptic flow 
$\ov{v}_{2,q}$.  Meson elliptic flow from the coalescence model 
is shown by open circles for a width of hadron Wigner function 
$\Delta_p=0.48$ GeV, by filled squares for $\Delta_p=0.24$ GeV, and
by filled circles for $\Delta_p=0.1$ GeV. Dashed lines are drawn to 
guide the eye.}
\label{v2scal-glob}
\end{figure}

Using the same quark distribution function of Ref.\cite{greco2}, 
which is essentially the blast wave model with isotropic radial flow
studied in Sect.\ref{blast}, we show in Fig. \ref{v2scal-glob} the 
result for mesons from the coalescence model based on a Monte 
Carlo evaluation of the coalescence integral. Three 
different values of $\Delta_p$ are used. The filled
squares are the result with $\Delta_p = 0.24$ GeV that was used 
in Ref.\cite{greco2,greco-res} to fit the experimental meson spectra 
at RHIC. In this case, the quark number scaling is violated by about 
$10\%$ in the scaled momentum region $p_T/n \simeq 1-2 $ GeV, as
already found in a previous work \cite{greco2,greco-res} and mentioned 
in the Introduction.  In Ref.\cite{moln-nscal}, a larger violation 
of about $20\%$ has been found in the quark number scaling of
hadron elliptic flow. This may be largely due to the fact that the 
corresponding value for $\Delta_p$ used in Ref.\cite{moln-nscal} is 
about a factor of two larger than our value due to the absence of 
the factor $9\pi/2$ in the hadron Wigner function.  Indeed, using 
$\Delta_p=0.48$ GeV in the present calculation leads to a much smaller 
scaled meson elliptic flow as shown by open circles, thus a much 
stronger violation of the quark number scaling of the meson elliptic 
flow.  In the scaled momentum region $p_T/n \simeq 1-2 $ GeV, the 
violation of the quark number scaling is consistent with the $20\%$ 
effect seen in Ref.\cite{moln-nscal}. Also shown in 
Fig. \ref{v2scal-glob} by filled circles are results 
obtained with a smaller $\Delta_p=0.1$ MeV. It is seen that they are
now very close to the quark elliptic flow given by the solid line,
as expected in the limit of $\Delta_p\to 0$.

\begin{figure}[ht]
\includegraphics[height=2.7in,width=3.5in]{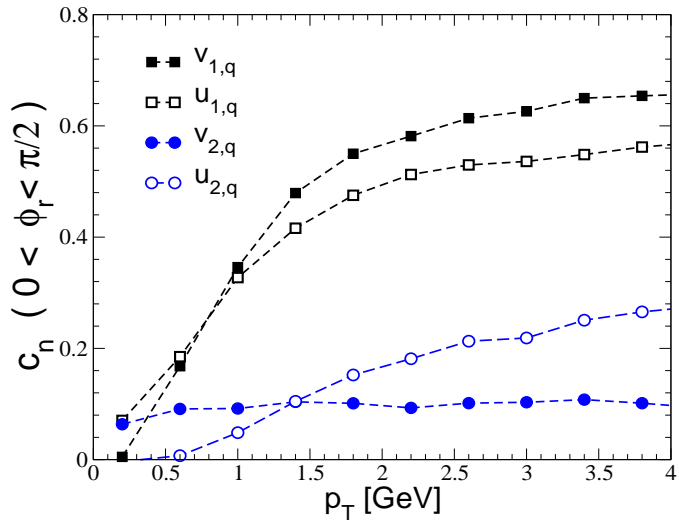}
\caption{Local harmonics of quark momentum distributions as a
function of transverse momentum from averaging over the first 
quadrant of the transverse plane.}
\label{cnq-therm}
\end{figure}

We have also evaluated the local harmonics of the momentum
distribution of quarks in the first quadrant. They are shown in 
Fig. \ref{cnq-therm} for $v_{1,q}$ (filled squares), $u_{1,q}$ 
(open squares), $v_{2,q}$ (filled circles), and $u_{2,q}$ 
(open circles). Both the real and imaginary local directed flows 
$v_{1,q}$ and $u_{1,q}$ are large, as in the anisotropic blast 
wave model of Sect.\ref{ablast} shown in Fig.\ref{v2qm-blast}. Our
results thus demonstrate that violation of the quark number 
scaling of hadron elliptic flow in the realistic model of 
Ref.\cite{greco2,greco-res} is not due to the large local anisotropic flow 
but is rather a result of the finite dispersion of quark momentum 
inside hadrons.

\section{summary}

To understand the observed quark number scaling in the elliptic flow 
of identified hadrons at RHIC, we have studied a number of models 
that range from the schematic one with a simple quark distribution
to the realistic blast wave model that is consistent with the underlying 
hydrodynamic and transport models. In the naive coalescence model, in
which only quarks with same momentum can coalescence, all models, 
except the blast wave model with an anisotropic radial flow, lead to 
scaling of hadron elliptic flow according to the constituent 
quark number. In the anisotropic blast wave model, the violation of this 
scaling is due to the azimuthal asymmetry in the quark coalescence 
probability as a result of the azimuthal angle dependence of the 
radial flow. In all cases, large local anisotropic flows are present, 
but they do not play any role in the final hadron elliptic flow. We 
have further demonstrated in a blast wave model that the quark number 
scaling can also be violated in a more realistic coalescence model 
which takes into account the momentum dispersion of quarks in hadron 
wave functions. Since effects due to anisotropic radial flow and 
hadron wave functions are small if realistic values are used for 
these quantities, observation of the quark number scaling of hadron 
elliptic flow in experiments thus provides a unique signature for 
hadronization via the quark coalescence as well as for the existence 
of the quark-gluon plasma prior the production of hadrons. 

\begin{acknowledgments}
We thank  Lie-Wen Chen for helpful communications and Hendrik van 
Hees and Ralf Rapp for useful conversations. This paper was based on 
work supported in part by the US National Science Foundation under Grant
Nos. PHY-0098805 and PHY-0457265 and the Welch Foundation under 
Grant No. A-1358.
\end{acknowledgments}

\end{document}